\documentclass[oneside]{article}
\usepackage{amsfonts}
\usepackage{stmaryrd}

\def\N{\ensuremath{\mathcal{N}}}

\def\Q{\ensuremath{\mathcal{Q}}}
\def\R{\ensuremath{\mathcal{R}}}

\def\Z{\ensuremath{\mathcal{Z}}}
\newcommand{\set}[1]{\{#1\}}
\def\ord{\ensuremath{\set{0,\ldots, n-1}}}
\def\trois{\ensuremath{\set{0,1,2}}}
\newcommand{\sm}{\ensuremath{\mbox{{\it sum}}}}
\newcommand{\lfp}{\ensuremath{\mbox{{\it lfp}}}}
\newtheorem{theorem}{Theorem}{}
\newtheorem{lemma}{Lemma}{}
\newtheorem{corollary}{Corollary}{}
{}
\newtheorem{definition}{Definition}{}
\newtheorem{example}{Example}{}
\newcommand{\F}{\ensuremath{(F,=)}}
\newcommand{\ra}{\rightarrow}

\newcommand{\cpp}{{\sc C++}}

\title{Logic programming beyond Prolog}
\author{M.H. van Emden\\
        Department of Computer Science\\
        University of Victoria \\
        Research report DCS-355-IR
       }
\date{}

\begin{document}
\maketitle
\date{}

\begin{abstract}
A logic program is an executable specification.  For
example, merge sort in pure Prolog is a logical formula, yet shows
creditable performance on long linked lists.  But such executable
specifications are a compromise: the logic is distorted by algorithmic
considerations, yet only indirectly executable via an abstract
machine.

This paper introduces {\sl relational programming},
a method that solves the difficulty with logic programming
by a separation of concerns.
It requires three texts: (1) the axioms, a logical formula
that specifies the problem and is not compromised by algorithmic
considerations, (2) the theorem, a logical formula that expresses
the idea of the
algorithm and follows from the axioms, and (3) the code,
a transcription of the theorem to a procedural language.
Correctness of the code relies on the logical
relationship of the theorem with the axioms
and relies on an accurate transcription of the theorem
to the procedural language.

Sorting is an example where relational programming has the advantage
of a higher degree of abstractness: the data to be sorted can be
any data type in C++ (the procedural language we use in our examples)
that satisfies the axioms of linear order,
while the pure-Prolog version is limited to data structures
in the form of linked cells.
We show another advantage of relational programs: they have a
model-theoretic and fixpoint semantics equivalent to each other and
analogous to those of pure Prolog programs.

\end{abstract}

\section{Introduction}\label{sec:intro}

We review some advantages and disadvantages of logic programming,
discuss how {\sl Elements of Programming} \cite{stpmcj09}
addresses one of the disadvantages,
and introduce {\sl relational programming} as a way of combining
the advantages of logic programming with those of
{\sl Elements of Programming}.

\subsection{Logic programming}

An advantage of logic programming
is that programs can be declaratively read
as definitions in logic of relations
yet can often be executed in Prolog.
Prolog is a versatile programming language
of adequate performance for a variety of applications.
See Bratko \cite{brtk01} for a sampling of applications
in artificial intelligence.
Another advantage is that the formal semantics of logic programs
can be defined in three ways: model-theoretic,
operational, and according to the fixpoint method
and that these can be shown to agree \cite{vEKow76,lld87}.

This paper is motivated
by a disadvantage and a disappointment of logic programming,
both exemplified by the use of pure Prolog for sorting.
The disadvantage is that in Prolog the only kind of sequence
that can be sorted is a data structure
in the form of linked cells.
This is the consequence of the fact that
in logic programming data are terms of logic.
While these are a surprisingly versatile data structure,
one might want to sort arrays, for example.

The disappointment has to do with program verification.
The fact that a logic program is a text that can be executed
as it is written and that is also a definition in logic
might lead to the expectation that
the executable text can serve as its own verification.
In this way logic programming
would eliminate the verification problem.

However, this is too optimistic,
as one can see when one wants to use Prolog for
sorting a list.
In many situations one can get satisfactory performance
with the program in Figure~\ref{prog:sortMrg}.
But this is not acceptable as a specification.
Acceptable as a specification
would be the program in Figure~\ref{prog:sortSpec},
which would even run as a Prolog program,
though it would take an amount of time
in the order of $n!$ for lists of length $n$.
\begin{figure}
\begin{center}
\begin{minipage}[t]{3in}
\hrule \vspace{2mm}
\begin{verbatim}
sort(V,W) :-
  split(V, V0, V1),
  sort(V0, W0), sort(V1, W1),
  merge(W0, W1, W).
\end{verbatim}
\hrule
\end{minipage}
\end{center}
\caption{
\label{prog:sortMrg}
A Prolog program for sorting.
It is assumed that suitable definitions of
{\tt split} and {\tt merge} have been added.
It behaves like merge sort,
but is it an acceptable specification?
}
\end{figure}

\begin{figure}
\begin{center}
\begin{minipage}[t]{3in}
\hrule \vspace{2mm}
\begin{verbatim}
sort(V, W) :- permutation(V, W), ordered(W).
\end{verbatim}
\hrule
\end{minipage}
\end{center}
\caption{
\label{prog:sortSpec}
A Prolog program for sorting.
This version directly reflects the specification:
the output is the sorted version of the input list
if it is an ordered permutation of it.
It runs as a Prolog program when supplemented with
suitable definitions for {\tt permutation} and {\tt ordered}.
}
\end{figure}

This example illustrates that not all definitions
in logic are equally suited as specification.
If proposed as a specification,
the program in Figure~\ref{prog:sortMrg} has to be rejected
as being distorted by algorithmic considerations.
A solution to this problem has appeared in
{\sl Elements of Programming} by Alexander Stepanov and Paul McJones
\cite{stpmcj09}.

\subsection{``Elements of Programming''}

In {\sl Elements of Programming}
Stepanov and McJones
derive many state-of-the-art algorithms in C++
by a method that uses
separate formulas of logic for the specification
and for the expression of the algorithm.
The first formula is referred to as ``axioms'';
the second as ``theorem''\footnote{
In what sense the theorem is justified by the axioms
is addressed in Section~\ref{sec:thmAxms}.
}.
Because of this separation,
the axioms can be pure in the sense of being free
of algorithmic considerations.
The role of the theorem is to express the idea of the algorithm.
That theorems written in logic can express algorithms
is familiar in logic programming.
This insight evolved independently in {\sl Elements of Programming}. 

According to this method, which we will refer to as EOP,
the axioms would contain a definition of the sortedness
relation, say, as being an ordered permutation.
But they would also contain the axioms for linear order.
As a result the theorem is true of any algebraic structure
that satisfies the axioms.
This is valuable for programming, as the axioms also
cover sequences other than Prolog lists;
for example arrays and, more generally, iterators \cite{strstrp13}.

In EOP the Prolog program for merge sort can take the
place of the theorem. 
In EOP the preferred programming language is not Prolog.
It is therefore necessary to transcribe the theorem to
code in the preferred language.
The result could be a program that sorts arrays,
among other possibilities for the data structure.

The following table compares EOP and logic programming.

\vspace{3mm}

\begin{tabular}{c||c|c}
& EOP & logic programming \\
\hline
\hline
Specification &
Figure~\ref{prog:sortSpec} plus
axioms &
Figure~\ref{prog:sortSpec} \\
& for linear order & \\
\hline
Theorem &
Prolog merge sort &
Prolog merge sort \\
\hline
Code & Figure~\ref{prog:sortCPP} &
Prolog merge sort \\
\end{tabular}

\begin{figure}
\begin{center}
\begin{minipage}[t]{4in}
\hrule \vspace{2mm}
\begin{verbatim}
  1 typedef char T;
  2 class Seg { // segment of an array
  3 public:
  4   T* bgn; unsigned n;
  5   Seg() {}
  6   Seg(T* bgn, unsigned n): bgn(bgn), n(n) {}
  7   void copy(Seg& w) {
  8     for (unsigned i=0; i<n; ++i)
          *(bgn+i) = *((w.bgn)+i);
  9   }
 10 };
 11 void merge(Seg& w0, Seg& w1, Seg& w) {
...
 28 }
 29 void split(Seg& v, Seg& v0, Seg& v1) {
...
 33 }
 34 void sort(Seg& v, Seg& w) {
 35   if ((v.n) <= 1) { w.copy(v); return; }
 36         // sort([], []) and sort([x], [x])
 37   Seg v0, v1; split(v, v0, v1);
 38         // split(v, v0, v1)
 39   T a[v0.n], b[v1.n];
 40   Seg w0(a, v0.n), w1(b, v1.n);
 41         // arrange local storage
 42   sort(v0, w0); sort(v1, w1);
 43         // sort(v0, w0), sort(v1, w1)
 44   merge(w0, w1, w);
 45         // merge(w0, w1, w)
 46 }
\end{verbatim}
\hrule
\end{minipage}
\end{center}
\caption{
\label{prog:sortCPP}
The Prolog program in Figure~\ref{prog:sortMrg} transcribed
to C++.
The comments in lines 34 -- 46
indicate the origins in the logic theorem.
}
\end{figure}
\paragraph{}
Advantages of EOP include the following.
\begin{enumerate}
\item
The relation to be computed is axiomatized
without algorithmic considerations.
These are relegated to the theorem.
\item
Not only the relation to be computed is axiomatized,
but also the data space.
For the latter many standard axiomatizations are available
in algebra textbooks: linear orders, partial orders, semigroups,
monoids, Archimedean monoids, $\ldots$ \cite{stpmcj09}.
\item
The abstractness of logic is exploited more fully than
in logic programming:
the C++ code can implement any algebraic structure
that satisfies the axioms.
As Hilbert is said to have remarked
in connection with his axioms for geometry:
instead of points, lines, and planes,
one can think of tables, beer mugs, and chairs.
\end{enumerate}

\subsection{Relational programming}

EOP has made an important contribution to
solving the problem of connecting specification
to code.
In this paper we introduce
{\sl relational programming}.

\begin{enumerate}
\item
{\bf Goal}\\[1mm]
To combine the advantages of logic programming and EOP.
\item
{\bf Method}\\[1mm]
Logic programming,
as introduced by Kowalski \cite{kow73,kow74},
is a package of two components:
(1) procedural interpretation of logic, and
(2) choice of language in which to express procedures.
The choice of procedural language made by Kowalski was (pure) Prolog.
To us EOP suggests procedural programming languages
other than Prolog.

In the context of EOP,
the right place to insert logic programming is the Theorem.
In this way we are less constrained than in logic programming:
instead of a formula executable in Prolog,
we only aim at one that is easily transcribed
to the procedural language of choice.

Because we are no longer tied to Prolog,
we can consider alternatives to clausal form in logic syntax.
We define relational programs as formulas reminiscent of
the if-halves of the Clark completion \cite{clrk78}
of a logic program.
These can be more directly transcribed to
a conventional procedural language
than Prolog as typically written.

\item
{\bf Results}\\[1mm]
We show that relational programs
can be given a model-theoretic and a fixpoint semantics
in the same manner as in \cite{vEKow76},
provided that Herbrand interpretations
are replaced by interpretations with
a fixed interpretation for the function symbols
over a freely chosen universe of discourse.

In principle, logic provides a high degree of abstractness.
For example, many structures satisfy the axioms for linear order.
When an algorithm is expressed as a relational program
that is a theorem\footnote
{
Not in the standard sense of logic.
The standard sense requires the theorem to be true in all
models of the axioms.
In relational programming it is only required that the minimal
model of the theorem satisfies the axioms.
See Section~\ref{sec:thmAxms}.
}
with respect to these axioms,
its transcriptions are correct with respect to any C++
classes or templates for sequences as long as they
conform to the axioms.

Our experience suggests that the transcription of a certain class
of relational programs to C++ is a routine task
that can be reliably executed.
A problem in coding is that the same function
can be written in many different ways.
One of the guidelines in software engineering
is to suppress this variability
and to write code in as stereotyped a fashion as possible.
Though this is widely accepted as a guideline,
there seems to be no agreement which stereotype to choose
from the many candidates.
Stereotyping by restricting C++ to 
transcription of relational programs
may prove to be a welcome contribution
to software engineering\footnote
{
It should be noted that Prolog also allows many
different variations for the same programming task.
Relational programs remove
much of this counterproductive variability.
}.

\item
{\bf Limitations}\\[1mm]
Logic programs can be nondeterministic and/or reversible.
These possibilities are lost in translation
to a conventional procedural language.
In a logic program the same parameter can be used for
input and for output.
Such programs are not suitable for transcription to C++.

Using Prolog ensures that there is no
discrepancy between the logic program
and what is executed.
In relational programming the possibility of error
is introduced by the transcription of the relational
program to conventional code.

\end{enumerate}

\section{Notation and basic notions}
\label{sec:notTerm}

Not all texts agree on the notations
and terminology in set theory and logic
that we need in this paper.
Therefore we collect in this section the necessary material,
terminology, and notation.

\subsection{Sets, functions, and relations}

\subsubsection{Sets}
We use \N, \Z, \Q\, and \R\ for the sets of natural numbers,
integers, rational numbers, and reals, respectively.
For $n \in \N$ we often need the set
$\{0, \ldots n-1\}$.
It is convenient to denote this set as $n$,
so that one can write, e.g., ``for all $i\in n$''
instead of the usual circumlocution.

\subsubsection{Functions}\label{sec:functions}
The set of functions that take arguments in a set $S$
and have values in a set $T$ is denoted $S \to T$.
This set is said to be the {\sl type} of a function
$f \in (S \to T)$.
We write $f(a)$ for the element of $T$
that is the value of $f$ for argument $a \in S$.

Suppose we have
$f \in S \to T$
and
$g \in T \to U$.
Then the composition $g \circ f$ of $f$ and $g$
is the function $h \in S \to U$ defined by
$x \mapsto g(f(x))$.

\subsubsection{Tuples}\label{sec:tuples}
We regard an $n$-tuple over a set $D$ as an object
$d = (d_0,\ldots,d_{n-1})$
in which an element of $D$ is associated
with each of the indexes $0,\ldots,n-1$.
It is convenient to view such a $d$ as a function
of type $n \to D$.
This formulation allows us to consider tuples
of which the index set is a set other than $\ord$.
Hence we define a tuple
as an element of the function set $I \to D$,
where $I$ is an arbitrary countable set to serve as index set.
$I \to D$ is the {\sl type} of the tuple.
When $t \in I \rightarrow D$ is regarded as a tuple,
we often write $t_i$ instead of $t(i)$ when $i\in I$.

\begin{example}
If $t$ is a tuple in $\set{x,y,z} \to \R$,
then we may have $t_x = 1.1$, $t_y = 1.21$,
and $t_z = 1.331$.
A more compact notation would be welcome;
we use
$t =
\begin{tabular}{c|c|c}
$x$ & $y$ & $z$ \\
\hline
$1.1$ & $1.21$ & $1.331$
\end{tabular}
$,
where the order of columns is immaterial.
\end{example}

\begin{example}
$t \in \trois \to \set{a,b,c}$,
where 
$t =
\begin{tabular}{c|c|c}
$2$ & $1$ & $0$ \\
\hline
$c$ & $c$ & $b$
\end{tabular}
$.
In cases like this, where the index set is of the form $\ord$,
we use the compact notation $t = (b,c,c)$,
using the conventional order of the index set.
\end{example}

\begin{example}
$(x_0, \ldots, x_{n-1})$ is a tuple $x$ of type
$n \rightarrow \{x_0, \ldots, x_{n-1}\}$.
\end{example}

\begin{example}
Suppose we have tuple $t \in n \rightarrow D$
for some set $D$ and
$(x_0, \ldots, x_{n-1}) \in (n \rightarrow \{v_0, \ldots, v_{n-1}\})$.
In the absence of repeated elements in $(x_0, \ldots, x_{n-1})$,
the inverse function $(x_0, \ldots, x_{n-1})^{-1}$ exists.
If we set $A = t \circ (x_0, \ldots, x_{n-1})^{-1}$,
then we have
$
(A(x_0), \ldots, A(x_{n-1})) = (t_0, \ldots, t_{n-1}) = t.
$
\end{example}

\subsubsection{Relations}
A relation is a set of tuples with the same type.
This type is the {\sl type} of the relation.
If $J \rightarrow D$ is the type of the relation,
then $J$ is the index set of the relation and $D$
is the domain of the relation.

Being subsets of $J \rightarrow D$,
relations of that type are partially ordered by set
inclusion.

A relation of type $n \rightarrow D$
is commonly denoted as $D \times \cdots \times D$ or as $D^n$.
This is an example of a relation consisting
of tuples indexed by numbers.
In general this is not the case.

\begin{example}
$\sm = \set{(x,y,z) \in (\trois \to \R) \mid x+y=z}$
is a relation of type $\trois \to \R$.
Compare this relation to the relation
$\sigma = \set{s \in (\set{x,y,z} \to \R) \mid s_x+s_y=s_z}$.
As their types are different,
they are different relations;
$(2,2,4) \in \sm$ is not the same tuple
as $s \in \sigma$ where
$s =
\begin{tabular}{c|c|c}
$x$ & $y$ & $z$ \\
\hline
$2$ & $2$ & $4$
\end{tabular}
$.
\hfill$\Box$
\end{example}

\subsection{Logic}\label{sec:logic}

A {\it signature} $L$ consists of
\begin{enumerate}
\item
A set of {\it constant symbols}.
\item
Sets of $n$-ary predicate symbols for
nonnegative integers $n$.
These include ``$=$'' for $n=2$,
and {\sl true} and {\sl false} for $n=0$.
We denote the arity of a predicate symbol $q$ by $|q|$.
\item
Sets of $n$-ary function symbols for
positive integers $n$.
We denote the arity of a function symbol $f$ by $|f|$.
\end{enumerate}

The language of logical formulas is determined by
a signature enhanced with a set $V$ of {\sl variables}.

To avoid notational minuti\ae\
we give the syntax in an abstract form.

A {\sl term} is a variable, a constant symbol,
or a pair consisting of a $k$-ary function symbol
and a tuple of $k$ terms.

An {\sl atomic formula} (or {\sl atom})
is a pair consisting of a $k$-ary predicate symbol
and a tuple of $k$ terms.
The index set of this tuple is $\{0,\ldots,k-1\}$.

A {\sl conjunction} ({\sl disjunction})is a tuple $C$
consisting of a set of formulas and an indication
that $C$ is a conjunction (disjunction).

An {\sl implication} is a pair of formulas
consisting of conclusion and a condition
together with an indication that the pair is an implication.

An {\sl existential} ({\sl universal}) quantification
is an expression $E$ consisting of a variable and a formula,
together with an indication that $E$
is an existential (universal) quantification.


\section{Relational programming}\label{sec:relProg}

Relational programming modifies logic programming
in the following ways.
\begin{enumerate}
\item
Programs are formulas according to Definition~\ref{def:relProg}
rather than sets of Horn clauses.
\item
Model-theoretic semantics is defined in terms of
the semantics of first-order predicate logic.
\item
The model-intersection property is expressed
in terms of $\F$-interpretations,
a generalization of Herbrand interpretations.
\item
Fixpoint semantics
is given in terms of a mapping
of the set of $\F$-interpretations to itself
using the semantics of formulas of first-order predicate logic.
\end{enumerate}

\subsection{General form of relational programs}

\begin{definition}\label{def:relProg}
A relational program is a sentence of the form
$$
\bigwedge_{q \in Q}
\left[\forall \left[A_q \leftarrow \bigvee_{r\in R_q} \exists
   \bigwedge_{s\in S_{qr}} B_{qrs}
              \right]
\right]
$$
where $A_q$ and $B_{qrs}$ are atomic formulas
and where quantification is over all free variables.
For all $q\in Q$,
$A_q$ and $[\bigvee_{r\in R_q} \exists
   \bigwedge_{s\in S_{qr}} B_{qrs}]$
have the same set of free variables.
For all $q\in Q$,
the arguments of $A_q$ are a sequence of variables
without any repetition.
\hfill$\Box$
\end{definition}

It will be convenient to abbreviate the expression in
Definition~\ref{def:relProg} to $A \leftarrow B$,
which stands for
$$\bigwedge_{q \in Q} \forall A_q \leftarrow B_q.$$
The procedural interpretation of relational programs is
shown in the following table.

\begin{center}
\begin{tabular}{l|r|l}
&
$\bigwedge_{q \in Q} \forall A_q \leftarrow B_q$
   & $A_q$ is the procedure header shared by\\
&   & alternative procedure bodies
\\[3mm]
\hline
&\\[1mm]
$B_q$ & $\bigvee_{r\in R_q} B_{qr}$
  & disjunction of alternative\\
&  & procedure bodies
\\[3mm]
\hline
&\\[1mm]
$B_{qr}$ & $\exists \bigwedge_{s\in S_{qr}} B_{qrs}$
  & procedure body: conjunction\\
&  & of atomic formulas
\end{tabular}
\end{center}

\begin{example}

Consider the relational program

\begin{eqnarray*}
\forall x.\; even(x) &\leftarrow&
   x = 0 \;\vee\; (\exists y.\; x = s(y) \wedge odd(y))\\
\forall x.\; odd(x) &\leftarrow&
   x = s(0) \;\vee\; (\exists y.\; x = s(y) \wedge even(y))
\end{eqnarray*}

This conforms to Definition~\ref{def:relProg} as shown in the
table below.
\end{example}

\begin{center}
\begin{tabular}{c|c|c}
$B_{qr}$ & $r=0$ & $r=1$                           \\[1mm]
\hline
\hline\\[-2mm]
$q = even$ & \parbox{0.75in}{$S_{qr} = \{0\}$ \\
                     $x=0$
                    }
           & \parbox{1.5in}{$S_{qr} = \{0,1\}$ \\
                     $\exists y.\; \underbrace{x=s(y)}_{S_{qr0}}
                            \wedge \underbrace{odd(y)}_{S_{qr1}}$
                    }                                   \\[3mm]
\hline\\[-2mm]
$q = odd$ & \parbox{0.75in}{$S_{qr} = \{0\}$ \\
                     $x=s(0)$
                    }
           & \parbox{1.5in}{$S_{qr} = \{0,1\}$ \\
                     $\exists y.\; \underbrace{x=s(y)}_{S_{qr0}}
                            \wedge \underbrace{even(y)}_{S_{qr1}}$
                    }                                   \\[3mm]
\end{tabular}
\end{center}

\begin{example}
The relational program that replaces the Prolog program
in Figure~\ref{prog:sortMrg} is
\begin{eqnarray*}
\forall v,w. sort(v,w) && \leftarrow\\
&& (v = nil \wedge w = nil) \vee \\
&& (\exists v_0,v_1,w_0,w_1. split(v, v_0, v_1) \wedge \\
&& sort(v_0,w_0) \wedge sort(v_1,w_1) \wedge \\
&& merge(w_0,w_1,w)\\
&& )
\end{eqnarray*}
\end{example}

\subsection{De Bruijn's algorithm}
Multiplication of integers can be done by repeated addition.
This is a slow process unless one makes use of the opportunities
to halve the multiplier in conjunction with doubling the 
multiplicand. 
This method has been recorded in the Rhind papyrus,
an ancient Egyptian document.
Similarly, division by repeated subtraction is a slow
process unless a similar trick is used.
That such a trick is available is less widely known.
It may have first appeared
in print in \cite{dkstr72}, where Dijkstra attributes
the algorithm to N.G. de Bruijn.

De Bruijn's algorithm is one of the many examples
where Stepanov and McJones \cite{stpmcj09} derive
executable code from a declarative statement concerning
a mathematical structure.
They exploit the abstractness of axiomatic characterizations
to make the algorithm applicable to structures other than natural numbers.
In case of de Bruijn's algorithm a suitable structure
is the Archimedian monoid, examples of which include
the integers, the rational numbers,
the binary fractions $n/2^k$,
the ternary fractions $n/3^k$,
and the real numbers.
When $a$ is divided by $b$ with quotient $m$ and remainder $u$,
the arguments $a$, $b$, and $u$ are monoid elements
and $m$ is an integer.

An Archimedean monoid is an ordered additive monoid
($0$ as neutral element, $+$ as binary operation)
in which the Archimedean property holds.
This property takes different forms for different ordered algebras.
In the case of an additive monoid
we define the Archimedean property by the axiom
$\forall a,b. \exists m,u.\; q(a,b,m,u)$
where $q(a,b,m,u)$ stands for
$$
  (0 \leq a \wedge 0 < b) \rightarrow (\exists m,u.\;m\cdot b + u = a
                \wedge 0 \leq u < b).
$$
Here $m\cdot b$
stands for $\underbrace{b + \cdots + b}_{m \; times}$.

Although not in the format of a theorem,
the top two equations on page 82 in \cite{stpmcj09}
effectively state a theorem that holds in Archimedean monoids.
We reformulate these equations in two steps,
first informally and then formally as Theorem~\ref{thm:quotRem}.

These equations can be reformulated as follows.
Suppose that dividing $a$ by $b$ results in quotient $m$
and remainder $u$.
Suppose that dividing $a$ by $b+b$ results in quotient $n$
and remainder $v$.
Then we have


\begin{enumerate}
\item
if $a<b$, then $m=0$ and $u=a$\\
\item
if $b \leq a < b+b$, then $m=1$ and $u=a-b$\\
\item
if $b+b \leq a$ and $v<b$, then $m=2n$ and $u=v$\\
\item
if $b+b \leq a$ and $b \leq v$, then $m=2n+1$ and $u=v-b$.
\end{enumerate}

Re items 2 and 4:
for monoid elements $x$ and $y$, $x-y$ is only used when 
$y \leq x$ and is shorthand for the $z$ that exists such
that $y+z = x$. As $+$ is the only operation in the monoid,
we write $b+b$ rather than $2b$.
But $m$ and $n$ are integers so that we see expressions such
as $m=2n$ and $m=2n+1$ in items 3 and 4.

In the second step we formalize the above equations as follows.
\begin{theorem}\label{thm:quotRem}
Let $q(a,b,m,u)$ mean that dividing in an Archimedean monoid
$a$ by $b$ gives $m$ with remainder $u$.
We assume $0 \leq a$ and $0<b$.
Then we have\\

\parbox{5in}{
\begin{tabbing}
$\forall a,b,$\=$m,u.\; q(a,b,m,u) \leftarrow$ \\
      \>$[(a<b \wedge m=0 \wedge u=a) \; \vee $\\
      \>$(b\leq a \wedge a<b+b \wedge m=1 \wedge u=a-b) \; \vee $\\
      \>$(b+b \leq a\;$\=$\wedge \; q(a,b+b,n,v)$
                $\wedge \; aux(b,m,u,n,v))]$ \\
\end{tabbing}
}\\
where

\parbox{5in}{
\begin{tabbing}
$\forall b,m,$\=$u,n,v.\; aux(b,m,u,n,v) \leftarrow $\\
   \>$[$\=$(v<b \wedge m = 2n \wedge u=v) \vee$\\
   \>\> $(b \leq v \wedge m = 2n+1 \wedge u = v-b)]$
\end{tabbing}
}
\hfill$\Box$
\end{theorem}

The theorem is transcribed in C-style pseudocode in
Figure~\ref{prog:qrPsC}.
For the compilable and executed version
in C++ see Figure~\ref{prog:qrCPP}.

\begin{figure}
\begin{center}
\begin{minipage}[t]{4in}
\hrule \vspace{2mm}
\begin{verbatim}
bool q(a, b, m, u){
  assert(0 <= a && 0 < b);
  if (a < b) { m = 0; u = a; return true; }
  if (b <= a && a < b+b) {
    m = 1; u = a-b; return true;
  }
  if (b+b <= a) { loc n; loc v; // local variables
    return q(a, b+b, n, v) && aux(b, m, u, n, v);
  }
  return false;
}
bool aux(b, m, u, n, v){
  if (v < b) { m = 2*n; u = v; return true; }
  if (b <= v) { m = 2*n+1; u = v - b; return true; }
  return false;
}
\end{verbatim}
\hrule
\end{minipage}
\end{center}
\caption{
\label{prog:qrPsC}
Pseudocode in C style for quotient and remainder
in Archimedean monoids.
For a compilable and executed
C++ version see Figure~\ref{prog:qrCPP}.
}
\end{figure}

\section{Model-theoretic semantics of relational programs}
\label{sec:modThSem}

The semantics of logic programs are simplified
because the interpretations are restricted
to Herbrand interpretations.
For relational programs the interpretations
are $\F$-interpretations for any universe,
so the semantics of logic is needed in its full generality.
Widely used texts \cite{mndlsn64,shnfld67,ndrtn72,grzgrczk74} agree
on this semantics.
The latter two refer to the semantics as
originating with Tarski \cite{tarski33},
with more accessible versions in \cite{tarski35,tarski56}.

\subsection{Structures and interpretations}\label{sec:interpr}

A {\sl structure} consists of a universe $D$
(also referred to as ``domain''), which is a set,
and numerically-indexed relations and functions
over $D$.\\[4mm]
A structure $S$ with universe $D$ (also referred to as ``domain'')
is an $L$-structure whenever
\begin{itemize}
\item
each constant in $L$ is associated with an element of $D$,
\item
each predicate symbol $q$ in $L$ is associated with a relation
in $S$ of type $|q| \rightarrow D$,
\item
and
each function symbol $f$ in $L$ is associated with a function
in $S$ of type $D^{|f|} \rightarrow D$.
\end{itemize}

\begin{definition}\label{def:DF}
Let $L$ be a signature and let $D$ be the universe of an
$L$-structure.
Let $F$ be the set of function symbols of $L$.
A \F-set is a set of interpretations
with the following properties:
(1) have the same domain $D$,
(2) have the same interpretation for the function symbols in $F$,
and
(3) the binary predicate symbol ``$=$'' is mapped to the identity
on $D$.
\hfill$\Box$
\end{definition}
Thus the interpretations of a \F-set differ only in the
interpretations of the predicate symbols.

For a given $F$,
$\F$-interpretations only differ in the relations
that are the interpretations of the predicate symbols
other than `='.
Thus we can view an $\F$-interpretation $I$ as a
vector of relations indexed by the set $Q$
of predicate symbols.
The component $I_q$ of $I$
that is indexed by an $|q|$-ary predicate symbol $q$
is a relation of type $|q| \rightarrow D$. 

\begin{definition}\label{def:preceq}
Let $I_0$ and $I_1$ be $\F$-interpretations with the
same signature, the same mapping $F$ and the same universe.
We define $I_0 \preceq I_1$ to mean
that $[I_0]_q \subseteq [I_1]_q$ for all $q\in Q$,
where $Q$ is the set of predicate symbols in $L$.

We denote by $\sqcup S$ the least upper bound,
if it exists,
of a set $S$ of $\F$-interpretations.
\hfill$\Box$
\end{definition}
Note that $\preceq$ is a partial order.

\begin{example}
Let $L$ be the signature with constants $0$ and $1$
and with binary operators $+$ and $\times$.
$D$ consists of the natural numbers \N\
and $F$ maps the symbols to the usual functions over $D$.
With these parameters in place,
the $\F$-interpretation gives terms
the values that are conventional for arithmetic expressions.
\end{example}

\begin{example}
Let any signature $L$ be given.
The set of variable-free $L$-terms can be the universe of a
structure with signature $L$.
Let $F$ map every function symbol $f$ in $L$ to the
function with map
$$
(a_0,\ldots,a_{|f|-1}) \mapsto
f(a_0,\ldots,a_{|f|-1}).
$$
This $\F$-interpretation is the Herbrand interpretation for
signatures without predicate symbols.
\end{example}

\subsection{Semantics of formulas}\label{sec:semForm}

Consider a variable-free $L$-formula and an interpretation $I$
for it.
This interpretation will be the basis of the determination
of the meaning $M^I$ of variable-free terms and formulas of logic.

\begin{definition}\label{def:meaningVF}
Semantics of variable-free terms and formulas
under interpretation $I$ is defined as follows.
\begin{itemize}
\item
$M^I(c) = I(c)$ if $c$ is a constant.
\item
$M^I(f(t_0,\ldots,t_{n-1}))
=
(I(f))(M^I(t_0),\ldots,M^I(t_{n-1})))$
if $f$ is a function symbol.
\item
$q(t_0,\ldots,t_{k-1})$
is satisfied by I iff
$ (M^I(t_0),\ldots,M^I(t_{k-1}))
   \in I(q)
$
if $q$ is a predicate symbol.
\item
A {\sl conjunction}
$\set{F_0, \ldots, F_{n-1}}$
of formulas
is satisfied by $I$ iff
$F_i$ is satisfied by $I$ for all $i \in n$.
\item
A {\sl disjunction}
$\set{F_0, \ldots, F_{n-1}}$
of formulas
is satisfied by $I$ iff
$F_i$ is satisfied by $I$ for at least one
$i \in n$.
\end{itemize}
\hfill$\Box$
\end{definition}

We now consider meanings of formulas
with a set $V$ of free variables,
possibly, but not typically, empty.
Let $\alpha$ be an {\sl assignment},
which is a function in $V \rightarrow D$,
assigning an individual in $D$ to every variable.
In other words, $\alpha$ is a tuple of elements of $D$
indexed by $V$.
As meanings of expressions with variables
depend on $\alpha$, we write $M^I_\alpha$
for the function mapping a term to an element
of the universe $D$.
$M^I_\alpha(F)$ asserts that a formula $F$ with set $V$
of free variables is satisfied
with assignment $\alpha\in(V\ra D)$
by interpretation $I$ with domain $D$.

\begin{definition}\label{def:meaningFV}
$M^I_\alpha$ is defined as follows.
\begin{itemize}
\item
$M^I_\alpha(t) = \alpha(t)$ if $t$ is a variable
\item
$M^I_\alpha(c) = I(c)$ if $c$ is a constant
\item
$M^I_\alpha(f(t_0,\ldots,t_{n-1}))
   = (I(f))(M^I_\alpha(t_0),\ldots,M^I_\alpha(t_{n-1}))).
$
\item
$q(t_0,\ldots,t_{k-1})$
is satisfied by $I$ with $\alpha$ iff\\
$ (M^I_\alpha(t_0),\ldots,M^I_\alpha(t_{k-1}))
   \in I(q).
$
\item
A conjunction
$\set{F_0, \cdots, F_{n-1}}$
is satisfied by $I$ with $\alpha$
iff the formulas $F_i$
are satisfied by $I$ with $\alpha$, for all $i \in n$.
\item
A disjunction
$\set{F_0, \cdots, F_{n-1}}$
is satisfied by $I$ with $\alpha$
iff the formulas $F_i$
are satisfied by $I$ with $\alpha$, for at least one $i \in n$.
\item
If $F$ is a formula,
then $\exists x. F$ is satisfied by $I$ with $\alpha$
iff there is a $d \in D$ such that $F$
is satisfied by
$I$ with $\alpha_{x|d}$
where $\alpha_{x|d}$ is the assignment
that maps $x$ to $d$ and maps the other variables
according to $\alpha$.
\item
If $F$ is a formula,
then $\forall x. F$ is satisfied by $I$ with $\alpha$
iff for all $d \in D$,
$F$ is satisfied by $I$ with $\alpha_{x|d}$
where $\alpha_{x|d}$ is the assignment
that maps $x$ to $d$ and maps the other variables
according to $\alpha$.
\end{itemize}
\hfill$\Box$
\end{definition}

So far, $M^I$ has assigned meanings to variable-free terms.
This is now extended as follows to terms with variables.

\begin{definition} \label{def:funcSem}
If $t$ is a term with set $V$ of variables,
then
$M^I(t)$ is the function of type
$(V \to D) \to D$
that maps
$\alpha \in (V \to D)$ to $M^I_\alpha(t) \in D$.
\hfill$\Box$
\end{definition}

\begin{example}
Suppose that
$t$ is $s(s(x))$,
$t'$ is $x+2$,
$D = \Z$,
$I$ maps the function symbol $+$
to addition among the integers \Z\
and maps $s$ to the successor function.
Now $M^I(t)$ and $M^I(t')$ are the same function in
$(\set{x} \to \Z) \to \Z$.
\hfill$\Box$
\end{example}

\begin{example}
$t$ is $x+2\times y+3\times z$,
$D = \Z$,
$I$ maps
the function symbol $+$
to
addition among integers
and maps $\times$
to
multiplication.
Then $M^I(t)$ is the function of type 
$ (\set{x,y,z} \to \Z) \to \Z $
with map $ \alpha \mapsto M^I_\alpha(t) $
e.g.

$ (M^I(t))(
$
\begin{tabular}{c|c|c}
$x$ & $y$ & $z$ \\
\hline
$3$ & $2$ & $1$
\end{tabular}
$
) = 10$,
which is $ M_\alpha^I(t)$ with 
$\alpha =
\begin{tabular}{c|c|c}
$x$ & $y$ & $z$ \\
\hline
$3$ & $2$ & $1$
\end{tabular}.
$
\hfill$\Box$
\end{example}

The following definition follows
Tarski {\it et al.}
\cite{hmt71},
Cartwright \cite{crt84}, page 377,
and Clark \cite{clrk91}.
It does for formulas what Definition~\ref{def:funcSem} does for terms.
\begin{definition}\label{def:bareDenot}
Let $F$ be a formula with set $V$ of free variables.
We define
$$ M^I(F) = \{\alpha \in (V \to D) \mid
   F \mbox{ is satisfied by } I \mbox{ with } \alpha
   \}.
$$
\hfill$\Box$
\end{definition}

\noindent
Thus $M^I(F)$ is a relation of type $V \rightarrow D$.

\begin{example}
$M^I(x\times x + y \times y < 2 \wedge x > y) = $
   \set{
\begin{tabular}{c|c}
$x$ & $y$ \\
\hline
$1$ & $0$
\end{tabular}
}
where $D = \Z$ and $\times$, $+$, and $<$ have the
usual interpretations.
\hfill$\Box$
\end{example}

If $V$ is empty,
then Definition~\ref{def:bareDenot}
gives the semantics of a closed formula, a sentence.
This conforms to the conventional definition of satisfaction
if we identify the relation $\{\}$ with ``not satisfied by $I$''
and identify being satisfied by $I$ with the relation
that is the singleton set containing the empty tuple.
In fact, one may define logical implication
in terms of Definition~\ref{def:bareDenot}.
\begin{definition}\label{def:logImpl}
If $A$ and $T$ are sentences,
then $A \models T$ holds iff
for all interpretations $I$ we have
$M^I(A) \subseteq M^I(T)$.
\hfill$\Box$
\end{definition}

\subsection{Model-theoretic semantics of relational programs}

The mere fact that predicate symbols are to be
interpreted as relations,
combined with the fact we have a precisely defined
semantics of first-order predicate logic,
does not make it immediately obvious
how to use a sentence of logic as a definition of
the relations referred to in this sentence.
In \cite{vEKow76} this is done for logic programs.
The solution given in that paper is restricted to Herbrand interpretations.
In the absence of that crutch
we have to consider afresh the question:
\begin{quote}
{\sl
How do we use a sentence of logic to define
the relations named in the sentence?
}
\end{quote}

Formulas are connected to relations by the fact that their
predicate symbols are interpreted as relations.
Hence a plausible answer to the question is:
\begin{quote}
{\sl
A sentence defines a set of relations as the relations
in the interpretation that makes the sentence true.
}
\end{quote}
However, there may be more than one
such interpretation, or there may not be any.
So this approach does not work.
The less obvious approach in the remainder of this section does.
The reason that it does so is that it allows us to show
that there exists at least one interpretation
and that there is {\sl the least} interpretation
that makes the sentence true.
The relations in this least interpretation are,
by our definition,
the ones defined by the relational program.

To show that a relational program has a least \F-model
for given $F$, our starting point is the definition of
\F-model. As relational programs are sentences of logic,
the special case of an empty set of free variables of
Definition~\ref{def:bareDenot} applies.
This makes a new definition superfluous.
But to emphasize this point we do add the following.
\begin{definition}
An \F-interpretation $I$ for a relational program $P$
is a model of $P$ if $P$ is true in $I$.
\hfill$\Box$
\end{definition}

The following characterization will turn out to be useful.

\begin{lemma}\label{lem:inclusion}
An \F-interpretation $I$ is a model of a relational
program $A \leftarrow B$ with set $Q$ of predicate symbols
iff $M^I(A_q) \supseteq M^I(B_q)$ for all $q \in Q$. 
\hfill$\Box$
\end{lemma}

{\sl Proof}\\
(If)\\
Suppose $I$ is not a model of $P$
and assume that $M^I(A_q) \supseteq M^I(B_q)$.
\begin{tabbing}
MMMMMMMMMMMMMMMMMMMMMMMMMMM\= MM\= \kill
$I$ falsifies $\forall (A_q \leftarrow B_q)$ for some $q \in Q$ 
                         \> $\Rightarrow$ \> (1) \\
$\exists \alpha \in (V_q \rightarrow D). \: A_q \leftarrow B_q$
is not true in $I$ with $\alpha$
                         \> $\Rightarrow$ \> (2) \\
$\exists \alpha \in (V_q \rightarrow D). \:B_q$
is true and $A_q$ is false in $I$ with $\alpha$
                         \> $\Rightarrow$ \> (3) \\
$\exists \alpha \in (V_q \rightarrow D).\: \alpha \in M^I(B_q)$
and $\alpha \not\in M^I(A_q)$
                         \> $\Rightarrow$ \> (4) \\
$\exists q \in Q.\: M^I(A_q) \not\supseteq M^I(B_q)$.
\end{tabbing}
(1,2,3): Definition~\ref{def:meaningFV};
(4): Definition~\ref{def:bareDenot}.\\[2mm]
(Only if)\\
Assume $I$ is a model of $P$.
\begin{tabbing}
MMMMMMMMMMMMMMMM\= MM\= \kill
$\alpha \in M^I(B_q)$
                     \> $\Rightarrow$ \> (1) \\
$\exists r \in R_q.\: B_{qr}$ true in $I$ with $\alpha$
                     \> $\Rightarrow$ \> (2) \\
$A_q$ true in $I$ with $\alpha$
                     \> $\Rightarrow$ \> (3) \\
$\alpha \in M^I(A_q)$
\end{tabbing}
(1): Definition~\ref{def:bareDenot};
(2): assumption;
(3): Definition~\ref{def:bareDenot}.
\hfill$\Box$

Whether, and how, this allows a relational program
to define relations depends on the properties of the models.

\subsection{The model-intersection property}

\begin{lemma}\label{lem:atomic}
Let $L$ be a non-empty set of $\F$-interpretations as defined
in Section~\ref{sec:interpr}.
Let $q(t_0,\ldots,t_{|q|-1})$ be an atomic formula.
We have
$$
M^{\cap L}(q(t_0,\ldots,t_{|q|-1})) = 
\cap_{I \in L} M^I(q(t_0,\ldots,t_{|q|-1})).
$$
\hfill$\Box$
\end{lemma}
{\sl Proof}

Let $V$ be the set of variables in $q(t_0,\ldots,t_{|q|-1})$.
\begin{tabbing}
MMMMMMMMMMMMMMMMMMMMMMMMMMMM\= MM\= \kill
$M^{\cap L}(q(t_0,\ldots,t_{|q|-1})) $  \> $=$ \> (1) \\
$\{\alpha \in V\rightarrow D \mid q(t_0,\ldots,t_{|q|-1})
  \mbox{ true in } \cap L \mbox{ with } \alpha\}$  \> $=$ \> (2)\\
$\{\alpha \in V\rightarrow D \mid 
     (M^{\cap L}_\alpha(t_0),\ldots M^{\cap L}_\alpha(t_{|q|-1}))
     \in [\cap L]_q \}$  \> $=$ \> (3)\\
$\{\alpha \in V\rightarrow D \mid 
     (M_\alpha(t_0),\ldots M_\alpha(t_{|q|-1})  ) \in [\cap L]_q
                                     \}$  \> $=$ \> (4)\\
$\{\alpha \in V\rightarrow D \mid 
     \forall I \in L.\;
        (M_\alpha(t_0),\ldots M_\alpha(t_{|q|-1})  ) \in I_q
                                     \}$  \> $=$ \> (5)\\
$\{\alpha \in V\rightarrow D \mid 
     \forall I \in L.\;
       q(t_0,\ldots,t_{|q|-1}) \mbox{ true in } I \mbox { with } \alpha
                                     \}$  \> $=$ \> (6)\\
$ \cap_{I \in L}\{\alpha \in V\rightarrow D \mid 
       q(t_0,\ldots,t_{|q|-1}) \mbox{ true in } I \mbox { with } \alpha
                                     \}$  \> $=$ \> (7)\\
$\cap_{I \in L} M^I(q(t_0,\ldots,t_{|q|-1}))$
\end{tabbing}
(1) Definition~\ref{def:bareDenot},
(2) Definition~\ref{def:meaningFV},
(3) the fact that $L$ only contains $\F$-interpretations,
so that the meaning of terms is independent of the interpretation,
(4) the definition of $L$,
(5) Definition~\ref{def:meaningFV},
(6) the definition of $L$,
and
(7) is by Definition~\ref{def:bareDenot}.
\hfill$\Box$

\begin{theorem}\label{thm:leastModel}
If $L$ is a non-empty set of $\F$-models of $P$,
then $\cap L$ is an $\F$-model of $P$.
\hfill$\Box$
\end{theorem}

{\sl Proof.}
We assume that $\cap L$ is not a model and show that this
leads to a contradiction.
\begin{tabbing}
MMMMMMMMMMMMMMMMMMMMMMMMMMMM\=MM\= \kill
$\cap L$ is not a model \> $\Rightarrow$ \> (1)\\
$\cap L$ falsifies $A_q \leftarrow B_q$
  for at least one $q \in Q$
                        \> $\Rightarrow$ \> (2)\\
$M^{\cap L}(A_q) \not\supseteq M^{\cap L}(B_q)$
  for at least one $q \in Q$
       \> $\Rightarrow$ \> (3)\\
$\exists \alpha \in V_q \rightarrow D.\;$
  $\alpha \in M^{\cap L}(B_q)$ and
  $\alpha \not\in M^{\cap L}(A_q)$
       \> $\Rightarrow$ \> (4)\\
$\exists \alpha \in V_q \rightarrow D.\;$
  $\forall I \in L.\; \alpha \in M^I(B_q)$ and
  $\alpha \not\in M^I(A_q)$
       \> $\Rightarrow$ \> (5)\\
$\exists \alpha \in V_q \rightarrow D.\;$
  $\forall I \in L.\; \alpha \in M^I(A_q)$ and
  $\alpha \not\in M^I(A_q)$
\end{tabbing}

(4) is by the definition of $L$ and Lemma~\ref{lem:atomic},
and
(5) is by the assumption that $I$ is a model of $P$.
\hfill$\Box$

\section{Fixpoint semantics of relational programs}\label{sec:fixpSem}

Given a relational program $P$ of the form $A \leftarrow B$,
we use the vector $B$ of right-hand sides to define a map
$T_{P}$ from the set of $\F$-interpretations of $P$ to itself.
We plan to show that $I \supseteq T_{P}(I)$ has a unique
least solution and that this equals the least model of $P$.

At first sight it might seem that one can simply define
$T_{P}(I) = M^I(B)$.
However, the $q$-component of $M^I(B)$ is not an interpretation
for the predicate symbol $q$,
which is what the $q$-component of 
$T_{P}(I)$ has to be.

Consider one of the conjuncts of $P$:
$
q(x_0,\ldots, x_{|q|-1}) \leftarrow B_q.
$
$M^I(B_q)$ is a relation consisting of tuples $t$
indexed by the set 
$\{x_0,\ldots, x_{|q|-1}\}$.
The $q$-component of $T_{P}(I)$ is a relation
consisting of tuples indexed by the set $\{0,\ldots, |q|-1\}$.
As there are no repeated occurrences of a variable in
$q(x_0,\ldots, x_{|q|-1})$,
the inverse
$(x_0,\ldots, x_{|q|-1})^{-1}$
exists so that
$t \circ (x_0,\ldots, x_{|q|-1})^{-1}$
is a tuple indexed by 
$\{x_0,\ldots, x_{|q|-1}\}$.
This observation suggests the following definition.

\begin{definition}\label{def:fixpOp}
Let $P$ be a relational program of the form $A \leftarrow B$,
with $Q$ as set of predicate symbols.
For every $q \in Q$, let $A_q$ be $q(x_0,\ldots, x_{|q|-1})$.
We define $T_P$ as a map from the set of $\F$-interpretations
for $P$ to itself.
$T_{P}(I)$ is defined as
the vector of relations indexed by $Q$ that
has the $q$-component
$$ \{
t \in (|q| \ra D) \mid
B_q \mbox{ is true in } I \mbox{ with }
   t \circ (x_0, \ldots, x_{|q|-1})^{-1}
\}. $$
\hfill$\Box$
\end{definition}
When $T_P$ is meant to be computable,
one has to ensure that the functions of the \F-interpretations
are computable.

\begin{theorem}
For any relational program $P$
and $\F$-interpretations $I_0$ and $I_1$,
$I_0 \preceq I_1$
implies
$T_P(I_0) \preceq T_P(I_1)$.
That is, $T_P$ is monotonic.
\hfill$\Box$
\end{theorem}

{\sl Proof}\\
It suffices to show that
$[I_0]_q \subseteq [I_1]_q$
implies
$[T_P(I_0)]_q \subseteq [T_P(I_1)]_q$
for every $q \in Q$,
where $Q$ is the set of predicate symbols in $P$.

\begin{tabbing}
MMMMMMMMMMMMMMMMMMMMMMMMMMMMMMMMMMMMMMMM\=MMM\kill
$t \in [T_P(I_0)]_q$ \>  $\Rightarrow$ (1)\\
$B_q$ true in $I_0$ with $t \circ (x_0,\ldots,x_{|q|-1})^{-1}$
                     \>  $\Rightarrow$ (2)\\
$B_{qr}$ true in $I_0$ with $t \circ (x_0,\ldots,x_{|q|-1})^{-1}$
for at least one $r \in R_q$
                     \>  $\Rightarrow$ (3)\\
$B_{qrs}$ true in $I_0$ with $t \circ (x_0,\ldots,x_{|q|-1})^{-1}$
for at least one $r \in R_q$
and all $s \in S_{qr}$
                     \> $\Rightarrow$ (4) \\
$B_{qrs}$ true in $I_1$ with $t \circ (x_0,\ldots,x_{|q|-1})^{-1}$
for at least one $r \in R_q$
and all $s \in S_{qr}$
                     \> $\Rightarrow$ (5) \\
$B_{qr}$ true in $I_1$ with $t \circ (x_0,\ldots,x_{|q|-1})^{-1}$
for at least one $r \in R_q$
                     \> $\Rightarrow$ (6)\\
$B_q$ true in $I_1$ with $t \circ (x_0,\ldots,x_{|q|-1})^{-1}$
                     \> $\Rightarrow$ (7) \\
$t \in [T_P(I_1)]_q$
\end{tabbing}
\noindent
(1) Definition~\ref{def:fixpOp},
(2) $B_q$ is a disjunction,
(3) $B_{qr}$ is a conjunction,
(4) $I_0 \preceq I_1$,
(5) $B_{qr}$ is a conjunction,
(6) $B_q$ is a disjunction,
(7) Definition~\ref{def:fixpOp}.
\hfill$\Box$

\paragraph{}
From monotonicity we conclude (see \cite{lld87}):
\begin{corollary}
For any relational program $P$, any universe $D$,
and any mapping $F$ from function symbols to functions
over $D$, $T_P$ has a unique least fixpoint.
\end{corollary}
We write the least fixpoint of $T_P$ as $\lfp(T_P)$.

\subsection{Existence of (least) fixpoint}

Definition~\ref{def:fixpOp} is the way it is to make the
resulting $T$ play the same role as the $T$
in the fixpoint semantics of logic programs \cite{vEKow76,lld87}.
In this paper the key theorem (the one in Section 7 of that paper),
which holds for logic programs,
is shown in Theorem~\ref{thm:modEqvFP}
to hold for relational programs,
provided we generalize Herbrand interpretations to \F-interpretations.

\begin{theorem}\label{thm:modEqvFP}
Let $I$ be an $\F$-interpretation of a relational program $P$.
Then we have that $I$ is a model of $P$ iff
$T_P(I) \preceq I$.
\hfill$\Box$
\end{theorem}

{\sl Proof}\\
Let $Q$ be the set of predicate symbols in $P$.\\
(Only if) Assume $I$ is a model of $P$ and assume
$t$ is an element of the $q$-component of $T_{P}(I)$ for
some $q \in Q$.
Assume $A_q$ is $q(x_0,\ldots,x_{|q|-1}).$

\begin{tabbing}
MMMMMMMMMMMMMMMMMMMMMMMMMMM\=MMM\kill
$t \in [T_P(I)]_q $\> $\Rightarrow$ (1)\\
$B_q$ true $I$ with $t \circ (x_0,\ldots,x_{|q|-1})^{-1}$
    \> $\Rightarrow$ (2)\\
$A_q$ true $I$ with $t \circ (x_0,\ldots,x_{|q|-1})^{-1}$
    \> $\Rightarrow$ (3)\\
$q(x_0,\ldots,x_{|q|-1})$ true in $I$
   with $t \circ (x_0,\ldots,x_{|q|-1})^{-1}$
    \> $\Rightarrow$ (4)\\
$t \in I_q$
\end{tabbing}
(1): Definition~\ref{def:fixpOp}, (2): $I$ is a model,
(3): $A_q$ is the atom $q(x_0,\ldots,x_{|q|-1})$, and
(4): Definition~\ref{def:meaningFV}.

\paragraph{}
(If) Assume $T_{P}(I) \preceq I$ and assume that
$\alpha \in M^I(B_q)$ for some $q \in Q$.
\begin{tabbing}
MMMMMMMMMMMMMMMMMMMMMMMMMM\=MMM\kill
$\alpha \in M^I(B_q)$\> $\Rightarrow$ (1)\\
$B_q$ true in $I$ with $\alpha$ \> $\Rightarrow$ (2)\\
$B_q$ true in $I$ with $d \circ (x_0,\ldots,x_{|q|-1})^{-1}$
     \> $\Rightarrow$ (3)\\
$d \in [T_{P}(I)]_q$ \> $\Rightarrow$ (4)\\
$d \in I_q$ \> $\Rightarrow$ (5)\\
$q(x_0,\ldots,x_{|q|-1})$ true in $I$
   with $d \circ (x_0,\ldots,x_{|q|-1})^{-1}$
      \> $\Rightarrow$ (6)\\
$q(x_0,\ldots,x_{|q|-1})$ true in $I$ with $\alpha$
      \> $\Rightarrow$ (7)\\
$A_q$ true in $I$ with $\alpha$ \> $\Rightarrow$ (8)\\
$\alpha \in M^I(A_q)$
\end{tabbing}
(1) by Definition~\ref{def:bareDenot};
(2) let $d = \alpha\circ(x_0,\ldots,x_{|q|-1})$,
then we have $\alpha = d\circ(x_0,\ldots,x_{|q|-1})^{-1}$
because there are no repeated variables in
$(x_0,\ldots,x_{|q|-1})$;
(3) by Definition~\ref{def:fixpOp};
(4) by assumption $T_{P}(I) \preceq I$;
(5) Definition~\ref{def:meaningFV};
(6) using definition of $\alpha$;
(7) $A_q$ is the atom $q(x_0,\ldots,x_{|q|-1})$;
(8) by Definition~\ref{def:bareDenot}.

Hence $M^I(A_q) \supseteq M^I(B_q)$
for all $q \in Q$,
so that $I$ is a model, via Lemma~\ref{lem:inclusion}.
\hfill$\Box$

\subsection{Computational characterization of the minimal model}

\begin{lemma}\label{lem:fixpComp}
Let $P$ be a relational program and
let an \F-set of interpretations be given for it.
We have
$\sqcup\{T^n_P(\bot) \mid n \in \N\} = \lfp(T_P)$.
\hfill$\Box$
\end{lemma}
{\sl Proof}\\
We abbreviate $\sqcup\{T^n_P(\bot)\mid n\in\N\}$ by $L$.

We first show that $L \preceq lfp(T)$.
Consider
$T^n_P(\bot) \preceq \lfp(T_P)$ for all $n\in\N$,
which can be proved by induction on $n$:
the base case $n=0$ follows from the definition of $\bot$;
for the induction step we have that
$T_P^{n+1}(\bot) = T_P(T^n_P(\bot))
   \preceq T_P(\lfp(T_P)) = \lfp(T_P)$
using the monotonicity of $T_P$
and the definition $\lfp$ in addition to the induction assumption.

Thus we have that  
$T^n_P(\bot) \preceq \lfp(T_P)$ for all $n\in\N$.
In other words, $\lfp(T_P)$ is an upper bound of
$\{T^n_P(\bot) \mid n\in\N\}$.
It remains to be noted that
$\sqcup\{T^n_P(\bot) \mid n \in \N\}$
is the {\sl least} upper bound.

In this way we have shown that $L \preceq \lfp(T_P)$.

To prove the Lemma, it remains to be shown that
$\lfp(T_P) \preceq L$.
Suppose it can be shown that $L$ is a fixpoint of $T_P$;
that is, that $T_P(L) = L$.
Then it would follow that $\lfp(T_P) \preceq L$
because $\lfp(T_P)$ is the {\sl least} fixpoint.

\paragraph{}
To show that $T_P(L)=L$,
we first show that $L\preceq T_P(L)$, as follows.
\begin{tabbing}
MMMMMMMMMMMMMMM\=MMM\kill
$d\in L$
\> $\Rightarrow$ (1)\\
$d\in \sqcup\{T^n_P(\bot) \mid n \in \N\}$
\> $\Rightarrow$ (2)\\
$\exists n\in\N.\; d\in T^n_P(\bot)$
\> $\Rightarrow$ (3)\\
$\exists n\in\N.\; d\in T_P(T_P^{n-1}(\bot))$
\> $\Rightarrow$ (4)\\
$d\in T_P(L)$
\end{tabbing}
(1) definition of $L$;
(3) definition of the power of $T_P$;
(4) monotonicity of $T_P$.

To show that $L=T_P(L)$,
it remains to be shown that $T_P(L) \preceq L$.
For each $q\in Q$, we prove that $[T_P(L)]_q \subseteq L_q$.
\begin{tabbing}
MMMMMMMMMMMMMMMMMMMMMMMMMMMMMMMMMMMMMM\=MMM\kill
$d\in [T_P(L)]_q$
\> $\Rightarrow$ (1)\\
$B_q$ true in $L$ with $d\circ (x_0,\ldots,x_{|q|-1})^{-1}$
\> $\Rightarrow$ (2)\\
$\exists r\in R_q.\; B_{qr}$
   true in $L$ with $d\circ (x_0,\ldots,x_{|q|-1})^{-1}$
\> $\Rightarrow$ (3)\\
$\exists r\in R_q \forall s\in S_{qr}.\; B_{qrs}$
   true in $L$ with $d\circ (x_0,\ldots,x_{|q|-1})^{-1}$
\> $\Rightarrow$ (4)\\
$\exists r\in R_q \forall s\in S_{qr}
   \exists n_{qrs}\in\N.\; B_{qrs}$ true in $T_P^{n_{qrs}}(\bot)$
      with $d\circ (x_0,\ldots,x_{|q|-1})^{-1}$
\> $\Rightarrow$ (5)\\
$\exists r\in R_q \exists n_{qr}\in\N.\; B_{qr}$
   true in $T_P^{n_{qr}}(\bot)$
     with $d\circ (x_0,\ldots,x_{|q|-1})^{-1}$
\> $\Rightarrow$ (6)\\
$\exists n_q\in\N.\; B_q$
   true in $T_P^{n_q}(\bot)$
     with $d\circ (x_0,\ldots,x_{|q|-1})^{-1}$
\> $\Rightarrow$ (7)\\
   $\exists n_q\in\N.\; d\in [T(T_P^{n_q}(\bot))]_q$
\> $\Rightarrow$ (8)\\
   $\exists n_q\in\N.\; d\in [T_P^{n_q+1}(\bot)]_q$
\> $\Rightarrow$ (9)\\
   $d\in [L]_q$
\end{tabbing}
(1) Definition~\ref{def:fixpOp}
    and $\{x_0,\ldots,x_{|q|-1}\}$ are the free variables of $B_q$,
(2) $B_q$ is a disjunction,
(3) $B_{qr}$ is a conjunction, $R_q$ as in
    Definition~\ref{def:relProg},
(4) $B_{qrs}$ is atom, with $S_{qr}$ as in
    Definition~\ref{def:relProg},
(5) take $n_{qr}= \max\{n_{qrs}\mid s\in S_{qr}\}$
    because $B_{qr}$ is a conjunction,
(6) take $n_{q}\in\{n_{qr}\mid B_{qr}\}$
    because $B_q$ is a disjunction,
(7) Definition~\ref{def:fixpOp},
(8) meaning of $n_q+1$ as power,
(9) $L$ abbreviates $\sqcup\{T^n_P(\bot)\mid n\in\N\}$;
    the sequence $T^n_P(\bot)$ is monotonically increasing
    with $n$.
\hfill$\Box$

\begin{example}
Consider the set of \F-interpretations
where the domain $D$ is the set \Q\ of rationals
and where $F=\{+,*,/\}$ with the interpretations
that are customary in \Q.
Let $P$ be the relational program
\begin{eqnarray*}
\forall x.\; q(x)
&\leftarrow& x=1 \vee \\
&&           (\exists y.\; x=0.5*(y+2/y) \wedge q(y))
\end{eqnarray*}
Let $I$ be the \F-interpretation such that
$I_q = \{1,3/2,17/12,\ldots\}$;
that is, the least set that contains $1$ and is closed under
the function in $2^\Q\ra 2^\Q$ with map
$$
S \mapsto (S \cup \{0.5*(y+2/y) \mid y\in S\}).
$$
It is well-known in numerical
analysis\footnote{e.g. \cite{henrici64}, Example 4, page 76}
that for every $\epsilon >0$ there is an element of $I_q$
that differs from $\surd 2$ by less than $\epsilon$.

Note that $I_q$ is a fixpoint of $T_P$
and that $I_q=\sqcup \{T^n_P \mid n\in\N\}$.
Therefore, $I_q$ is the least fixpoint of $T_P$.

These conclusions are also valid when $D$ is the set \R\
of reals.
But in this case it is also true that
$I'$ with $I'_q = I_q \cup \{\surd 2\}$
is a fixpoint of $T_P$.
Clearly, $I'$ is not the least fixpoint.
\hfill$\Box$
\end{example}

Lemma~\ref{lem:fixpComp} allows us to give a computational
chararacterization of the minimal model.
\begin{theorem}
For every relational program $P$ its minimal model equals
$\sqcup\{T^n_P(\bot) \mid n \in \N\}$.
\hfill$\Box$
\end{theorem}
{\sl Proof}\\
According to Theorem~\ref{thm:leastModel}, $P$ has a least model.
According to Theorem~\ref{thm:modEqvFP} this is the least fixpoint.
According to Lemma~\ref{lem:fixpComp}
this equals $\sqcup\{T^n_P(\bot) \mid n \in \N\}$.
\hfill$\Box$

\section{Future work}

\begin{itemize}
%
\item
Transcription of relational programs to C++ is easy enough.
However, those who are oppressed
by the size and complexity of C++
might be interested in the language
resulting from eliminating everything
not needed for the transcription of relational programs.
\item
Conversely, formal logic was formed more than a century ago and has,
with few exceptions, only been used for theoretical purposes.
Even textbooks on abstract algebra give the axioms informally.
Logic lacks facilities for writing large formulas
in a structured fashion.
It may benefit from some of the structuring facilities
that allow programs of many thousands of lines
to be written in conventional programming languages.
\end{itemize}

\section{Conclusions}
\begin{itemize}
\item
The work reported here
suggests the following method of programming.
First, express an algorithm in the form of a relational
program $P$.
Second, determine a suitably general family of structures
to which the algorithm is applicable and
write a list $A$ of axioms characterizing this family
such that $A$ is true in the minimal model of $P$.
Finally, transcribe the relational program to
a program $P'$ in a suitable procedural language. 
The result is a program $P'$
that has property $A$ in a sense that is defined
in terms of the semantics of first-order predicate logic.
In the case of C++ as the procedural language,
$P'$ can be compiled to efficient code,
even though it is written in a style
that is unusual in current practice.
The compiler's optimization capabilities can take care of
the superficially apparent inefficiencies.

\item
Fixpoint and model-theoretic semantics of logic programs with respect
to Herbrand interpretations generalize to these semantics for
relational programs with respect to \F-interpretations.

\item
Kowalski's Procedural Interpretation of Logic
has not only procedurally interpreted Horn clauses,
but also limited the language
for expressing procedures to pure Prolog.
The work reported here gives a procedural interpretation,
not of Horn clauses, but of non-clausal formulas
that are in the format of a relational program.
We leave open the choice of procedural language;
we do not propose to replace Prolog,
but propose to expand the scope of logic programming.
\end{itemize}

\section{Acknowledgements}
Thanks to Philip Kelly,
Paul McJones, and Areski Nait-Abdallah for useful
discussions.
I am indebted to the reviewers for their insightful remarks
and for their suggestions for improvement. 
I am grateful to
the University of Victoria and the Natural Sciences and Engineering
Research Council of Canada for the facilities provided.


\appendix

\section{Appendix: Glossary}\label{sec:glossary}

\begin{tabular}{ll}
$\ra$    & set of functions; see Section~\ref{sec:functions}\\

$\sqcup$ & least upper bound; see Definition~\ref{def:preceq}\\

$\preceq$ & arguments are partially ordered;
     see Definition~\ref{def:preceq}\\

$^{-1}$ & inverse of function, especially useful for inverse
of tuple without repeated elements; see Section~\ref{sec:tuples}\\
\end{tabular}

\begin{tabular}{ll}

domain & Section~\ref{sec:interpr} \\

\F-interpretation & Definition~\ref{def:DF}\\

\F-set &  Definition~\ref{def:DF}\\

$M^I$ & Definition~\ref{def:meaningVF} for variable-free,
   Definitions
   \ref{def:funcSem} and
   \ref{def:bareDenot}
   for case with (free) variables\\

$M_\alpha^I$ & Definition~\ref{def:meaningFV}\\

relational program & Definition~\ref{def:relProg} \\

signature & Section~\ref{sec:logic}  \\

structure & Section~\ref{sec:interpr}  \\
 
universe & Section~\ref{sec:interpr} \\
\end{tabular}

\section{Appendix: Programs}

\begin{figure}
\begin{center}
\begin{minipage}[t]{4.5in}
\hrule \vspace{2mm}
\begin{verbatim}
  1 #include <stdio.h>
  2 #include <assert.h>
  3 
  4 typedef unsigned QT; // quotient type
  5 class AM { // Archimedean Monoid
  6 public:
  7   double val; //floating-point for Archimedean Monoid
  8   AM (): val(0) {}
  9   AM (double val): val(val) {}
 10   static AM zero() { return AM(0); }
 11   friend AM operator+(const AM& x, const AM& y)
 12     { return AM(x.val + y.val); }
 13   friend AM operator-(const AM& x, const AM& y)
 14     { return AM(x.val - y.val); }
 15   friend bool operator<(const AM& x, const AM& y)
 16     { return x.val < y.val; }
 17   friend bool operator<=(const AM& x, const AM& y)
 18     { return x.val <= y.val; }
 19 };
 20 bool aux(const AM& b, QT& m, AM& u,
 21          const QT& n, const AM& v){
 22   if (v < b) { m = 2*n; u = v; return true; }
 23   if (b <= v) { m = 2*n+1; u = v - b; return true; }
 24   return false;
 25 }
 26 bool q(const AM& a, const AM& b, QT& m, AM& u){
 27   assert(0 <= a && 0 < b);
 28   if (a < b) { m = 0; u = a.val; return true; }
 29   if (b <= a && a < b+b) {
 30     m = 1; u = a-b; return true;
 31   }
 32   if (b+b <= a) { QT n; AM v;
 33     return q(a, b+b, n, v) && aux(b, m, u, n, v);
 34   }
 35   return false;
 36 }
 37 
 38 int main() {
 39   AM a(1000000001.1), b(17), u; QT m;
 40   if (q(a, b, m, u)) {
 41     printf("%d %lf\n", m, u.val/b.val);
 42     printf("%lf\n", a.val/b.val);
 43   } else assert(false);
 44 }
\end{verbatim}
\hrule
\end{minipage}
\end{center}
\caption{
\label{prog:qrCPP}
The C++ program for quotient and remainder in Archimedean monoids
transcribed from Theorem~\ref{thm:quotRem}.
}
\end{figure}

\end{document}